\documentclass{appolb}
\usepackage{graphicx}
\usepackage{amssymb,amsmath,epsfig}
\usepackage{mathrsfs}
\usepackage[usenames,dvipsnames]{color}
\usepackage[bookmarks]{hyperref}
\usepackage[svgnames]{xcolor}
\usepackage{subfigure}

\newcommand{\dvol}{\mbox{dvol}}

\begin{document}
\title{Kinetic gas disks surrounding Schwarzschild black holes
}
\author{Carlos Gabarrete and Olivier Sarbach
\address{Instituto de F\'isica y Matem\'aticas,
Universidad Michoacana de San Nicol\'as de Hidalgo,
Edificio C-3, Ciudad Universitaria, 58040 Morelia, Michoac\'an, M\'exico.}}

\maketitle

\begin{abstract}
We describe stationary and axisymmetric gas configurations surrounding black holes. They consist of a collisionless relativistic kinetic gas of identical massive particles following bound orbits in a Schwarzschild exterior spacetime and are modeled by a one-particle distribution function which is the product of a function of the energy and a function of the orbital inclination associated with the particle's trajectory. The morphology of the resulting configuration is analyzed.
\end{abstract}

\date{\today}


\section{Introduction}
\label{Sec:Introduction}

In recent years, there has been interest in analyzing the properties of solutions to the Vlasov equation on a fixed, curved background spacetime. In particular, such an analysis has been performed for a Schwarzschild background with the aim of understanding the Bondi-Michel and Bondi-Hoyle-Littleton accretion models for a collisionless kinetic gas~\cite{pRoS17,aGetal2021,pMoA2021b}. Also, kinetic analogues of the perfect fluid ``Polish doughnuts" configurations are discussed in~\cite{cGoS2021b}. Similarly to their fluid counterparts, they describe stationary and axisymmetric disks around black holes, where the individual gas particles follow bound timelike geodesics in a Schwarzschild spacetime. In~\cite{cGoS2021b} these configurations are modeled by a one-particle distribution function (DF) depending only on the energy $E$, azimuthal $L_z$ and total angular momentum $L$ of the particles. Examples are given in which the DF is described by a generalized polytropic ansatz~\cite{eAhAaL16,eAhAaL19} depending only on $E$ and $L_z$. In this article, we provide additional examples where the DF is a function of $E$ and the inclination angle $i$ defined by $\cos i = L_z/L$. We analyze the behavior of the resulting particle density and compute the total number of particles of the gas cloud as a function of the free parameters in our ansatz.

\section{The model}
\label{Sec:Model}

We work in the Schwarzschild exterior spacetime, written in the usual coordinates $(t,r,\vartheta,\varphi)$, with metric\footnote{We use units in which the speed of light and the gravitational constant are one.}
\begin{equation}
\label{Eq:Smetric}
g := -N(r) dt^2 + \frac{dr^2}{N(r)} + r^2 \left( d\vartheta^2 + \sin^2\vartheta d\varphi^2 \right), \quad N(r) := 1-\frac{2M}{r} > 0,
\end{equation}
where $M > 0$ is the mass of the black hole. Since this spacetime is static and spherically symmetric, the particle's rest mass $m$ is conserved along with $E$, $L$ and $L_z$. In terms of the orthonormal tetrad $e_{\hat{0}} = N(r)^{-1/2}\partial_t$, $e_{\hat{1}} = N(r)^{1/2}\partial_r$, $e_{\hat{2}} = r^{-1}\partial_\vartheta$, $e_{\hat{3}} = (r\sin\vartheta)^{-1}\partial_\varphi$, the four-momentum of the particles can be parametrized as
$p = p^{\hat{\mu}} e_{\hat{\mu}}$ with (see~\cite[Eq. (58)]{aGetal2021}):
\begin{equation}
(p^{\hat{\mu}}) = \left( \frac{E}{\sqrt{N(r)}}, \epsilon_r \sqrt{\frac{E^2 - V_L(r)}{N(r)}},
\frac{\epsilon_\vartheta}{r}\sqrt{L^2 - \frac{L_z^2}{\sin^2\vartheta}}, \frac{L_z}{r\sin\vartheta} 
\right),
\label{Eq:OrthonormalBasis}
\end{equation}
where the signs $\epsilon_r = \pm 1$ and $\epsilon_\vartheta = \pm 1$ determine the direction of motion in the radial and polar directions, respectively, and $V_L(r) = N(r)(m^2 + L^2/r^2)$ is the effective potential for the radial motion.

A collisionless relativistic gas consisting of identical massive particles of mass $m$ trapped in $V_L$ is described by a DF which relaxes in time to a DF depending only on integrals of motion. This is due to phase mixing, see e.g.~\cite{pRoS18a,pRoS20} and references therein. Here we assume, in addition, that the final configuration is axisymmetric, which implies that the DF has the form
\begin{equation}
f(x, p) = F(E,L,L_z),
\label{Eq:OneParticleDistributionFunction}
\end{equation}
for some function $F$ which we shall specify shortly. The relevant spacetime observables are the particle current density vector field $J$ and the energy-momentum-stress tensor $T$ defined by
\begin{equation}
J_{\hat{\mu}}(x) := \int_{P_x^+(m)} f(x, p) p_{\hat{\mu}} \dvol_x(p), \quad
T_{\hat{\mu}\hat{\nu}} (x) := \int_{P_x^+(m)} f(x, p) p_{\hat{\mu}} p_{\hat{\nu}} \dvol_x(p),\quad
\label{Eq:JT}
\end{equation}
where $\dvol_x(p) = dp^{\hat{1}}\wedge dp^{\hat{2}}\wedge dp^{\hat{3}}/p^{\hat{0}}$ is the Lorentz-invariant volume form on the future mass hyperboloid  $P_x^+(m)$ of mass $m$ at $x$, see~\cite{rAcGoS2021} for details.

\begin{figure}[htb]
\centerline{
\subfigure{\includegraphics[scale=0.2]{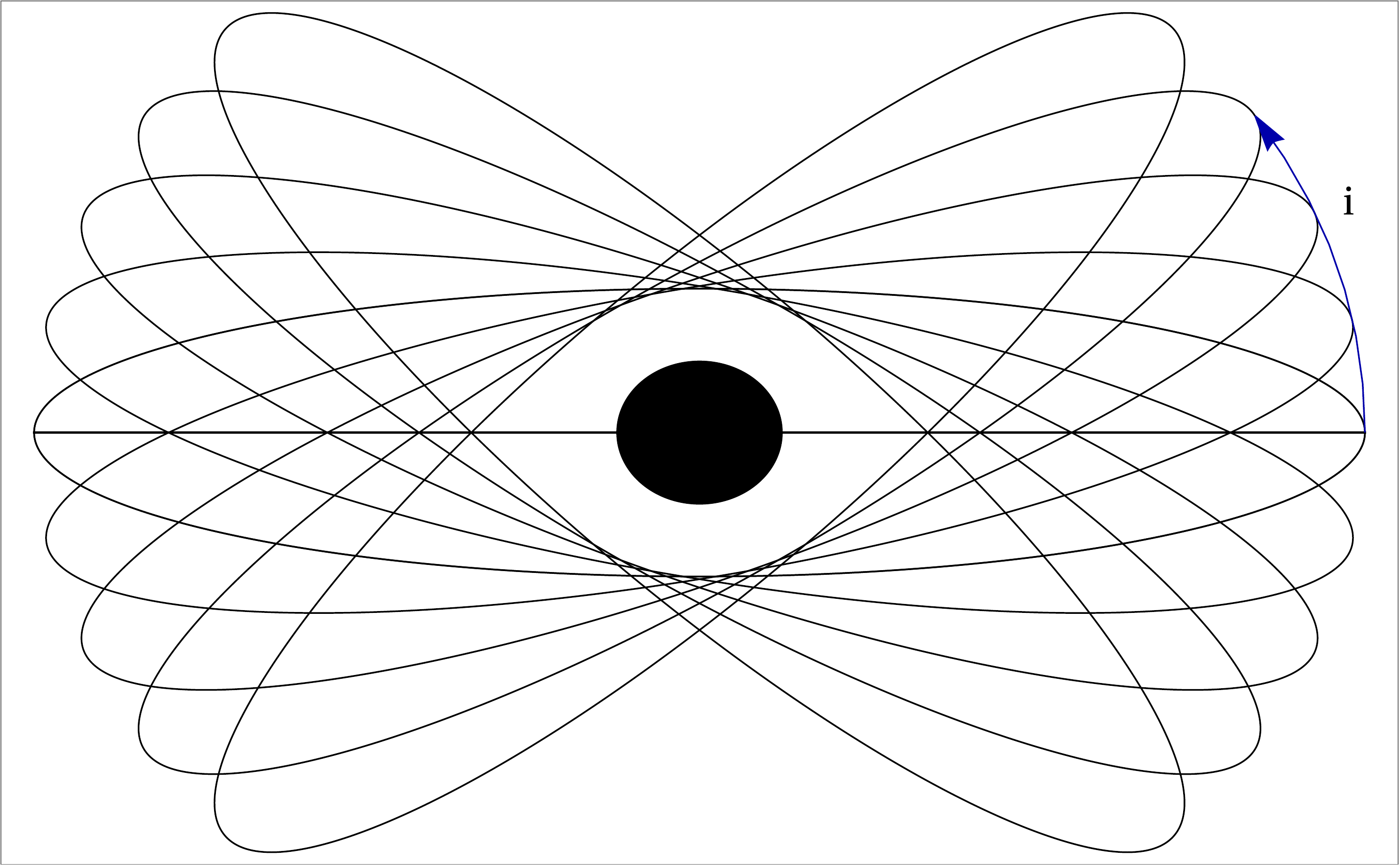}} 
\subfigure{\includegraphics[scale=0.2]{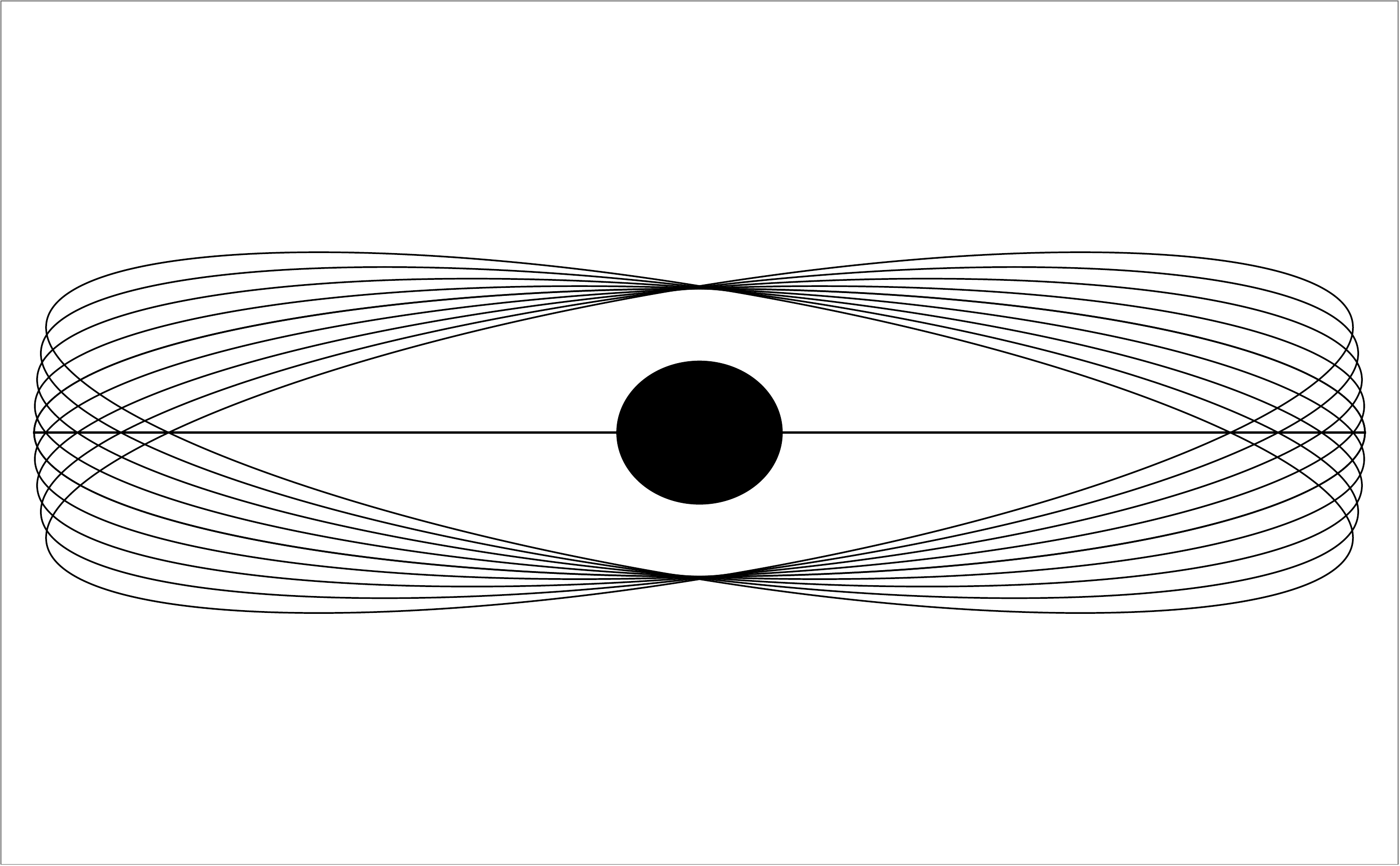}}} 
\caption{Illustration of the effect of the parameter $s$ (left panel for small $s$, right panel for large $s$). As $s$ increases, orbits confined to planes lying close to the equatorial one become more populated, such that the configuration becomes a thin disk in the limit $s\to \infty$.}
\label{Fig:InclinationAngle}
\end{figure}

For the following, we focus on the particular ansatz
\begin{equation}
\label{Eq:OneParticleDistrFunct}
F(E,L,L_z) := F_0(E) \cos^{2s}(i),\qquad
F_0(E) = \alpha \left(1 - \frac{E}{m} \right)_+^{k-\frac{3}{2}},
\end{equation}
where $\alpha > 0$, $k > 1/2$ are constants, $i$ is the inclination angle and $s\geq 0$ is a parameter. The notation $f_+$ refers to the positive part of the quantity $f$, that is $f_+ = f$ if $f > 0$ and $f_+ = 0$ otherwise. Here, the function $F_0$ is the general relativistic generalization of the polytropic ansatz~\cite{BT} while the parameter $s$ controls the concentration of the orbits near the equatorial plane $\vartheta = \pi/2$ (see Fig.~\ref{Fig:InclinationAngle}).
\\
For the following we introduce the dimensionless quantities $\xi := r/M$, $\lambda := L/(M m)$, $\varepsilon := E/m$, and $U_\lambda(\xi) := V_L(r)/m^2$ and parametrize the future mass hyperboloid $P_x^+(m)$ in terms of the quantities $(\varepsilon,\lambda,\chi)$, where the angle $\chi$ is defined by $(p^{\hat{2}},p^{\hat{3}}) = \frac{m\lambda}{\xi}(\cos\chi,\sin\chi)$ which implies $\cos i = \sin\vartheta \sin\chi$. For bound orbits, these quantities are restricted to the following domain (see~\cite[Appendix A]{pRoS16} and~\cite[Appendix A]{cGoS2021b}):
\begin{equation}
\label{Eq:Limits}
\varepsilon_{\text{c}}(\xi) < \varepsilon < 1, \quad \lambda_{\text{c}}(\varepsilon) \leq \lambda \leq \lambda_{\text{max}}(\varepsilon,\xi), \quad \hbox{and} \quad 0\leq \chi \leq 2\pi,
\end{equation}
where $\varepsilon_{\text{c}}(\xi)$ is the minimum energy at radius $\xi$, $\lambda_{\text{c}}(\varepsilon)$ is the critical value for the total angular momentum for which the maximum of the potential barrier in $U_\lambda(\xi)$ is exactly equal to $\varepsilon^2$, and $\lambda_{\text{max}}(\varepsilon,\xi)$ is the maximum angular momentum permitted at energy $\varepsilon$ and radius $\xi$. Note that the domain~(\ref{Eq:Limits}) is empty if $\xi < 4$, since for a Schwarzschild black hole the minimum radius for bound orbits is $r = 4M$.

For the ansatz~(\ref{Eq:OneParticleDistrFunct}) the fibre integrals in Eq.~(\ref{Eq:JT}) yield
\begin{equation}
\label{Eq:Jmu}
J_{\hat{\mu}}(x) = \frac{m^2\sin^{2s}\vartheta}{\xi^2}
\sum_{\epsilon_r = \pm 1}\int_{\varepsilon_{\text{c}}(\xi)}^1 \int_{\lambda_{\text{c}}(\varepsilon)}^{\lambda_{\text{max}}(\varepsilon,\xi)} 
\int_0^{2\pi} p_{\hat{\mu}} F_0(E) \sin^{2s}\chi \frac{d\varepsilon\; \lambda d\lambda\;d\chi}{\sqrt{\varepsilon^2 - U_\lambda(\xi)}},
\end{equation}
and similarly for $T_{\hat{\mu}\hat{\nu}}(x)$. Using the expressions~(\ref{Eq:OrthonormalBasis}) for the four-momentum, the non-vanishing orthonormal components of $J^{\hat{\mu}}$ and $T^{\hat{\mu}}{}_{\hat{\nu}}$ are
\begin{eqnarray}
\label{Eq:J0}
J^{\hat{0}} &=& 4\sqrt{\pi} \frac{\sin^{2s}\vartheta}{N^{3/2}} \frac{\Gamma(s+1/2)}{\Gamma(s+1)} m^3\int_{\varepsilon_{\text{c}}(\xi)}^1 d\varepsilon \varepsilon Y(\varepsilon,\xi)^{1/2}
F_0(m\varepsilon),\\
\label{Eq:T00}
T^{\hat{0}}{}_{\hat{0}} &=& -4\sqrt{\pi} \frac{\sin^{2s}\vartheta}{N^2} \frac{\Gamma(s+1/2)}{\Gamma(s+1)} m^4\int_{\varepsilon_{\text{c}}(\xi)}^1 d\varepsilon \varepsilon^2 
Y(\varepsilon,\xi)^{1/2}
F_0(m\varepsilon),\\
\label{Eq:T11}
T^{\hat{1}}{}_{\hat{1}} &=& \frac{4\sqrt{\pi}}{3} \frac{\sin^{2s}\vartheta}{N^2} \frac{\Gamma(s+1/2)}{\Gamma(s+1)} m^4\int_{\varepsilon_{\text{c}}(\xi)}^1 d\varepsilon 
Y(\varepsilon,\xi)^{3/2}
F_0(m\varepsilon),\\
\label{Eq:T22}
T^{\hat{2}}{}_{\hat{2}} &=& \frac{4\sqrt{\pi}}{3} \frac{\sin^{2s}\vartheta}{N^2} \frac{\Gamma(s+1/2)}{\Gamma(s+2)} m^4\int_{\varepsilon_{\text{c}}(\xi)}^1 d\varepsilon 
Y(\varepsilon,\xi)^{1/2}
Z(\varepsilon,\xi) F_0(m\varepsilon),
\qquad\\
\label{Eq:T33}
T^{\hat{3}}{}_{\hat{3}}  &=& (2s+1)T^{\hat{2}}{}_{\hat{2}},
\end{eqnarray}
where we have introduced the shorthand notation
\begin{equation}
Y(\varepsilon,\xi) := \varepsilon^2 - N(r)\left[1 + \frac{\lambda_c(\varepsilon)^2}{\xi^2} \right],\quad
Z(\varepsilon,\xi) := \varepsilon^2 - N(r)\left[1 - \frac{\lambda_c(\varepsilon)^2}{2\xi^2} \right].
\end{equation}
The quantities~(\ref{Eq:J0})--(\ref{Eq:T33}) determine the relevant macroscopic observables, namely the particle density $n = J^{\hat{0}}$, energy density $\mathcal{E} = -T^{\hat{0}}{}_{\hat{0}}$, and the principal pressures $P_1 = T^{\hat{1}}{}_{\hat{1}}$, $P_2 = T^{\hat{2}}{}_{\hat{2}}$, and $P_3 = (2s+1)P_2$. Note that all of these quantities have the dependency $\sin^{2s}\vartheta$ with respect to the polar angle $\vartheta$. In the limit $s = 0$ the configurations describe a spherical shell of gas trapped in the region $\xi > 4$, while for $s = 1/2,1,3/2,\ldots$ they are axisymmetric, the macroscopic variables being zero for $\xi \leq 4$ and along the axis $\vartheta = 0,\pi$. In the next section, we analyze the morphology of these configurations as a function of the parameters $k$ and $s$ for fixed total particle number.

\section{Total particle number and behavior of the particle density}
\label{Sec:Mass}

The (conserved) total particle number ${\cal N}$ is defined as minus the flux integral of the current density vector field with respect to a Cauchy surface. This in turn can be rewritten as an integral over the six-dimensional phase space parametrized by $(x^i,p_i)$. To compute this integral, it is convenient to transform $(x^i,p_i)$ to action-angle variables $(\mathcal{Q}^i,\mathcal{J}_i)$. The integral over the angle variables $\mathcal{Q}^i$ yields a factor $(2\pi)^3$ while the integral over the action variables can be rewritten in terms of the conserved quantities $(E,L,L_z)$, taking into account that $d^3\mathcal{J} = T(E,L) dE dL dL_z /2\pi$, where $T(E,L)$ is the period function for the radial motion. For the Schwarzschild spacetime this function can be expressed in terms of elliptic integrals and has the form $T(E,L) = 2 M \varepsilon \left[ \mathbb{H}_2 - \mathbb{H}_0 \right]$ (see~\cite[Appendix A]{pRoS18a} and~\cite{cGoS2021b} for the explicit form of $\mathbb{H}_2$ and $\mathbb{H}_0$ in the Schwarzschild case). For the ansatz~(\ref{Eq:OneParticleDistrFunct}) this yields the following expression for the total particle number:
\begin{equation}
\label{Eq:NumberParticles}
\mathcal{N} = \frac{16\pi^2}{2s+1}(M m)^3 \alpha \int_{\varepsilon_{\text{min}}}^1 d\varepsilon \; \varepsilon \left(1-\varepsilon \right)^{k-\frac{3}{2}}_+ 
\int_{\lambda_{\text{c}}(\varepsilon)}^{\lambda_{\text{ub}}(\varepsilon)} 
d\lambda\; \lambda \; (\mathbb{H}_2-\mathbb{H}_0),
\end{equation}
where $\varepsilon_{\text{min}} = \sqrt{8/9}$ and $\lambda_{\text{ub}}(\varepsilon)$ is given in~\cite[Appendix A]{pRoS16}. To compute this integral, it is convenient to re-parametrize the orbits in terms of their eccentricity $e$ and ``semi-latus rectum" $P$, related to the turning points $(\xi_1,\xi_2)$ by $\xi_1 = P/(1+e)$ and $\xi_2 = P/(1-e)$ and to the conserved quantities $(\varepsilon,\lambda)$ according to~\cite{wS02,jBmGtH15,pRoS18a,cGoS2021b} $(\varepsilon^2,\lambda^2) = ( P^{-1}[(P-2)^2 - 4e^2], P^2)/(P - e^2 -3)$. Here, $(P,e)$ are restricted to the domain $0 < e < 1$ and $P > 6 + 2e$. The resulting integral is then calculated numerically using MATHEMATICA. The total mass is simply $m{\cal N}$ and the total energy is given by the same expression as in Eq.~(\ref{Eq:NumberParticles}) with an extra factor $m\varepsilon$ inside the integral.

\begin{figure}[h!]
\centerline{
\subfigure{\includegraphics[scale=0.205]{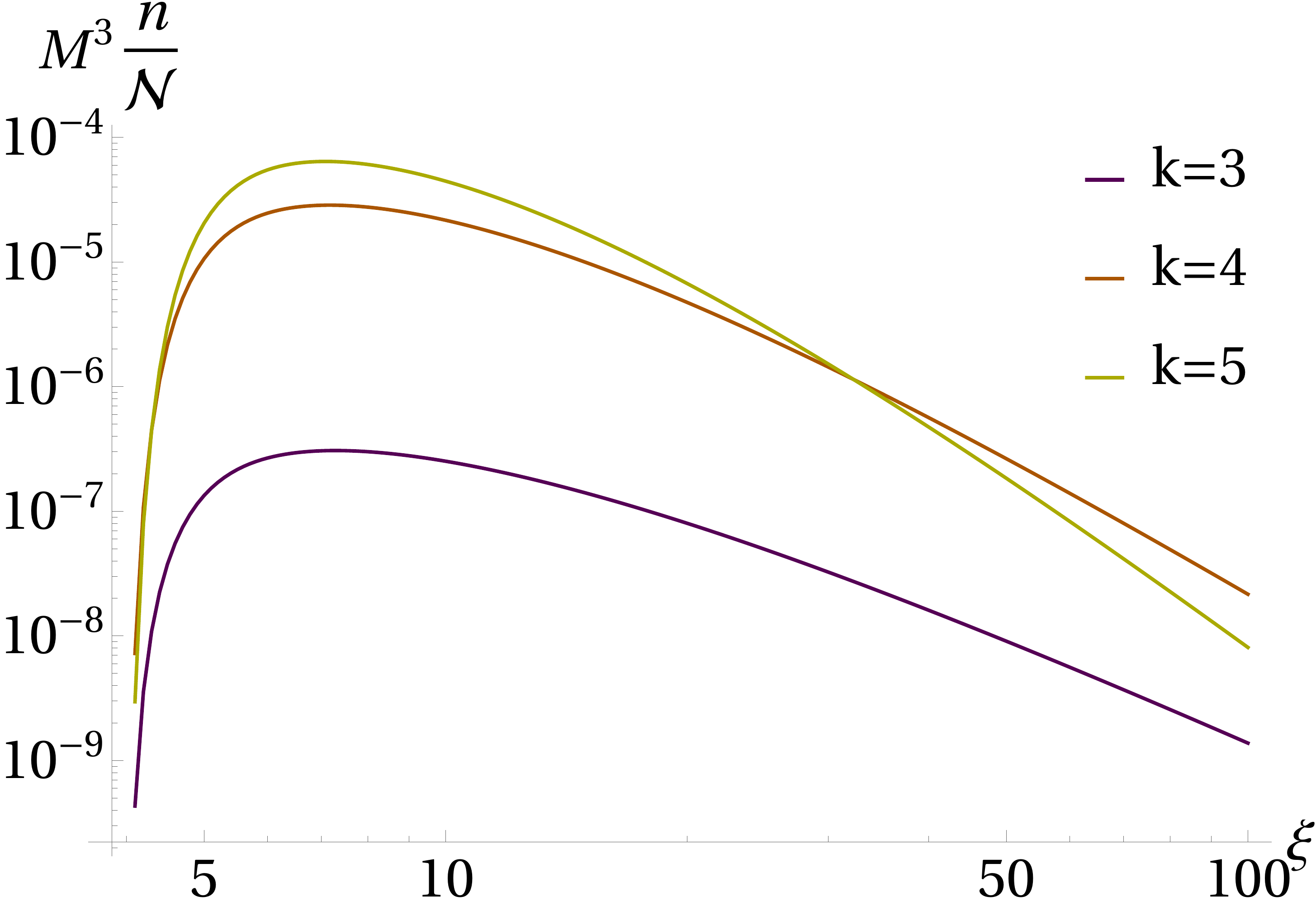}}
\subfigure{\includegraphics[scale=0.205]{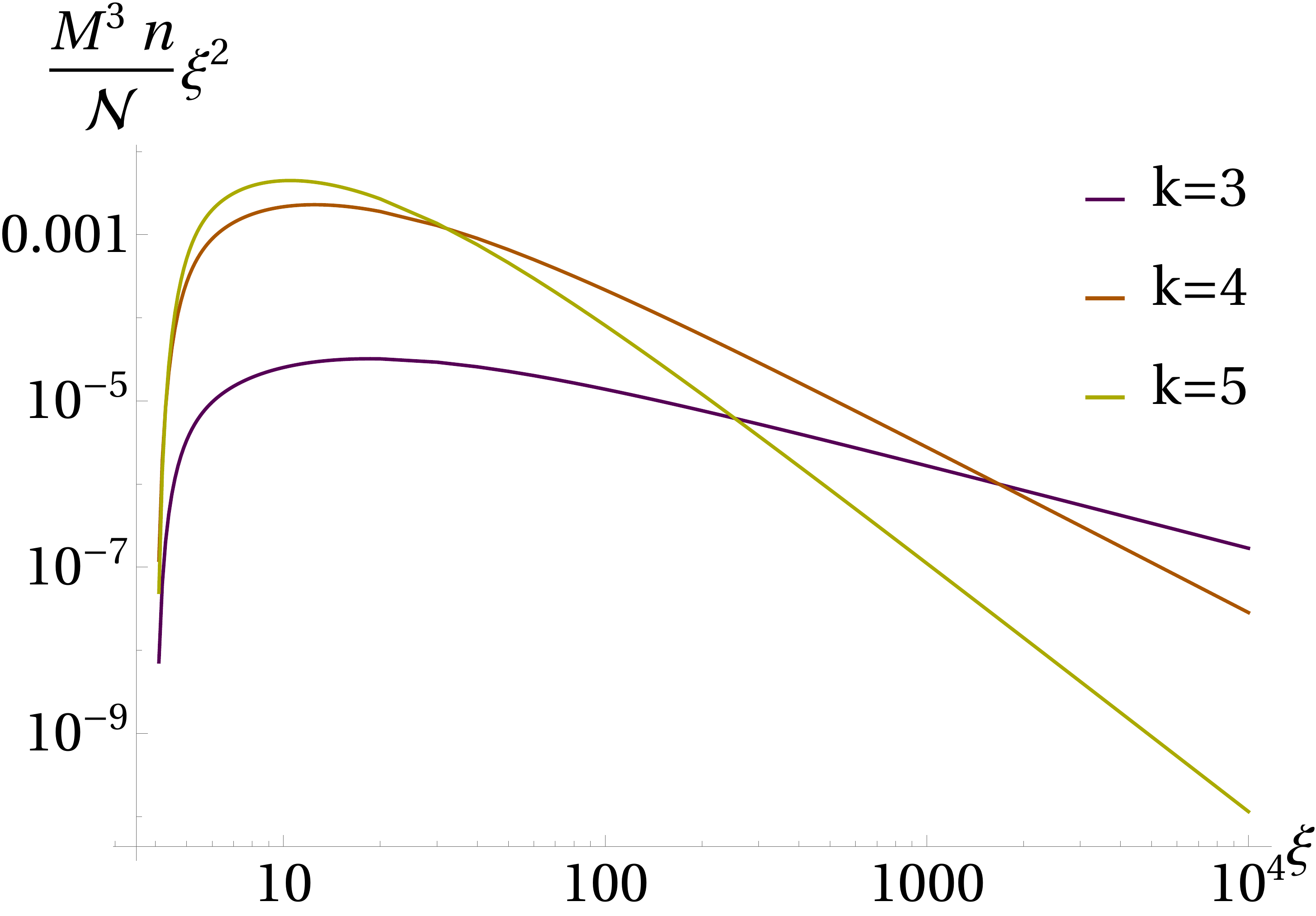}}}
\caption{Left panel: Dimensionless profile of the particle density in the equatorial plane for $k=3,4,5$ and $s=1$ in a logarithmic scale. Right panel: Same quantity multiplied with $\xi^2$ which shows that even though configurations with higher values of $k$ have a larger maximum, they have a faster decay at infinity.}
\label{Fig:ParticleDensitys1k345} 
\end{figure}
\begin{figure}
\centerline{
\subfigure{\includegraphics[scale=0.25]{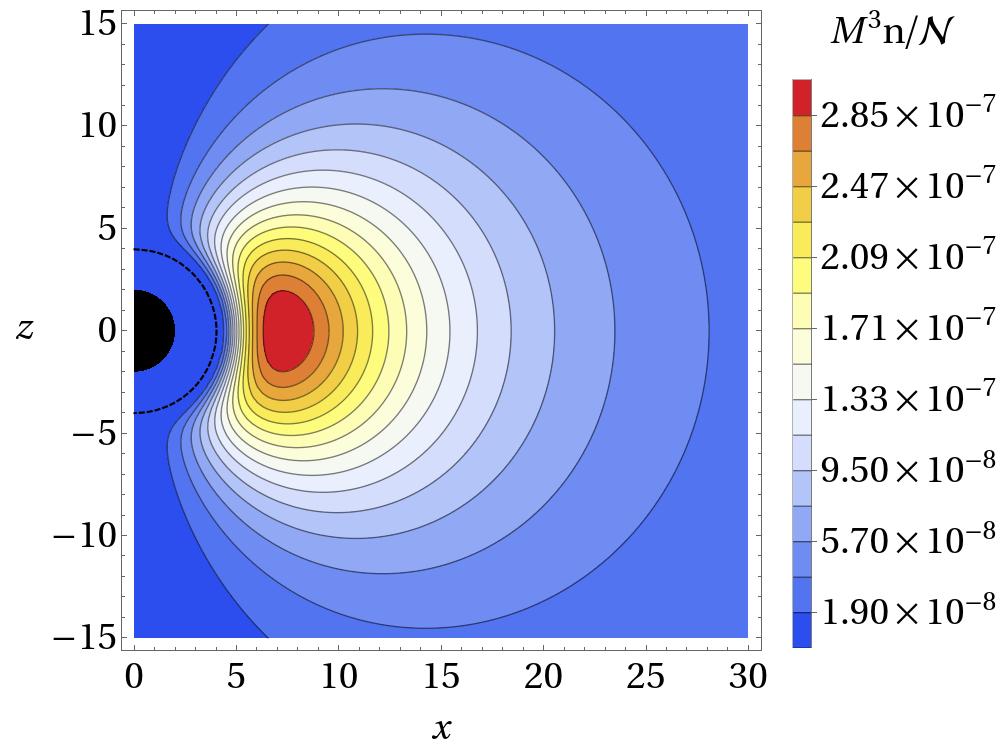}}
\subfigure{\includegraphics[scale=0.25]{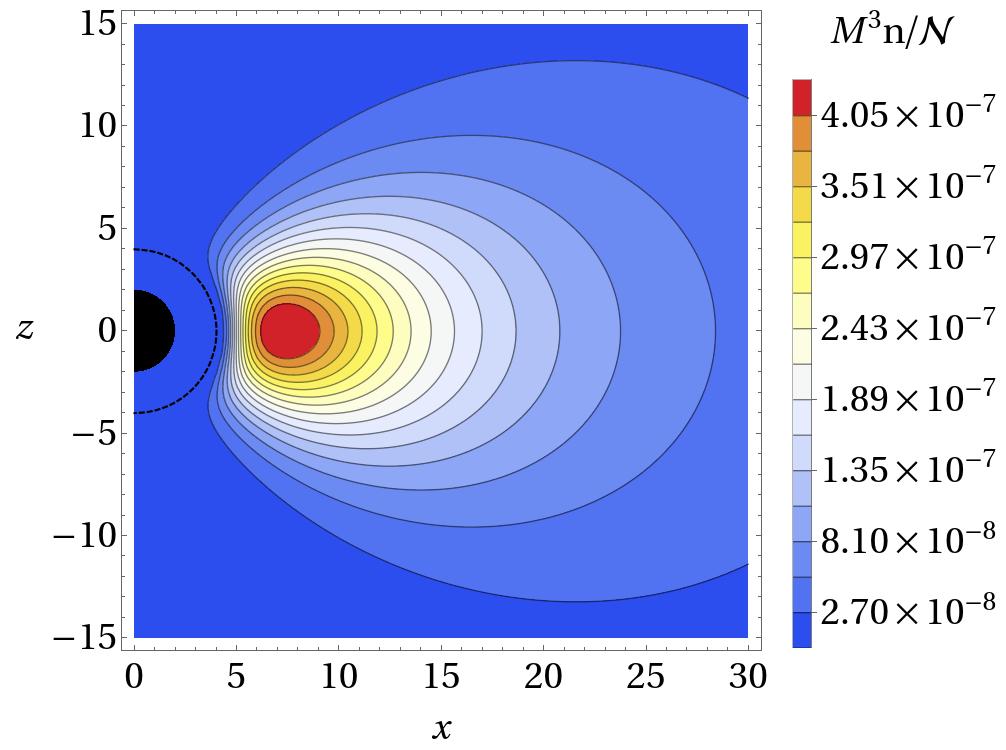}}}
\caption{Contour plots for the particle density in the $xz$-plane for the configurations with $k=3$ and $s=1$ (left panel) and $k=3$ and $s=3$ (right panel). Here, $(x,z) = r(\sin\vartheta,\cos\vartheta)$, the black region represents the black hole interior and the dashed black circles the interior boundary of the disk. As is visible from these plots, the configuration with higher $s$ yields a thinner disk.}
\label{Fig:ContourParticleDensitys1k3} 
\end{figure}

In Fig.~\ref{Fig:ParticleDensitys1k345} we show the dimensionless quantity $M^3 n/{\cal N}$ in the equatorial plane for several values of $k$ and $s=1$. In Fig.~\ref{Fig:ContourParticleDensitys1k3} we show contour plots of the same quantity in the $xz$-plane for $k=3$ and two different values of $s$.

\section{Conclusions}

We described a family of stationary and axisymmetric collisionless gas configurations which are trapped in the gravitational potential of a Schwarz-schild black hole. This family depends on two parameters $s$ and $k$ which control the thickness of the disk and its radial density distribution. An alternative model is discussed in detail in~\cite{cGoS2021b}. We expect these configurations to serve as a first approximation for the description of low-luminosity disks surrounding black holes.

We acknowledge support from a CIC Grant to Universidad Michoacana and CONACyT Frontier Project No. 376127.


\end{document}